\documentclass{ws-procs975x65}

\begin{document}

\title{Theoretical Description of kHz QPOs in accreting LMXB systems
}
\author{Banibrata Mukhopadhyay
\footnote{\uppercase{W}ork partially
supported by grant 80750 of the \uppercase{A}cademy of \uppercase{F}inland.}}
\address{Astronomy Division, P.O.Box 3000, University of Oulu,
FIN-90014, Finland; bm@cc.oulu.fi}
\author{Subharthi Ray
\footnote{\uppercase{P}resent address: \uppercase{IUCAA},
\uppercase{P}ost \uppercase{B}ag 4, \uppercase{G}aneshkhind,
\uppercase{P}une-411007, \uppercase{I}ndia; sray@iucaa.ernet.in}}
\address{Instituto de F\'isica, Universidade  Federal Fluminense,
Niter\'oi 24210-340 RJ, Brazil; sray@if.uff.br}
\author{Jishnu Dey\footnote{\uppercase{A}lso supported by \uppercase{SP/S2/K-03/01, DST, I}ndia.}}
\address{UGC Research professor, Department of Physics, Maulana Azad College, 8 Rafi Ahmed Kidwai Road,
Kolkata-700013; deyjm@giascl01.vsnl.net.in}
\author{Mira Dey\footnote{\uppercase{A}lso supported by \uppercase{SP/S2/K-03/01, DST, I}ndia.}}
\address{Department of Physics, Presidency College, 86/1 College Street, Kolkata-700073;
deyjm@giascl01.vsnl.net.in}

\maketitle

\abstracts{ We describe kHz QPOs from the hydrodynamical model of
accretion disks for LMXB systems. Out of the pair, the higher
frequency originates due to the viscous effects of an accretion
disk involving the formation of shock, while the lower one is due
to the Keplerian motion of accreting matter. Comparing our results
with observations for two fast rotating compact stellar
candidates, namely, 4U~1636$-$53 and KS~1731$-$260, we find that
they match to a very good approximation. }


One of the most important {\it questions} in present day
astrophysics is the origin of quasi-periodic oscillations (QPO)
observed from compact objects. Also the particular interests are
behind those which are of kilohertz (kHz) order and are supposed
to originate from neutron star candidates in the low mass X-ray
binary (LMXB) system. Interesting point is that those kHz QPO
frequencies are observed in a pair for a particular source and
difference of those pair frequency is almost same or half of the
spin frequency of the compact star. Moreover sometimes there is
the existence of side-band(s) of those QPOs. Since their
discovery, several (theoretical) models have been proposed to
describe their origin (e.g. see references\cite{to99,w03}) but to our knowledge
none of them could {\it completely} explain all of the related
phenomena.

Here we propose one such model to describe a few aspects of QPO
from the hydrodynamical description of accretion disks. As the
QPOs originate from close to the compact object, an inner edge of
the accretion disk exists in that region but in the sub-Keplerian
regime, because the gravitational attraction of compact objects
essentially wins over any other forces at that extreme inner
region.

Recently, a number  of  pseudo-Newtonian potentials have been
proposed which describe the  time varying  as well as
steady-state relativistic properties of accretion disks  around
rotating compact objects with hard surface (e.g., neutron star but
not black hole)\cite{mm03,mg03}. Therefore, here we like to
describe our disk in pseudo-Newtonian manner
incorporating the relativistic properties approximately.
A detailed discussions about the model equation set we need
to solve for accretion disk structure, are given in Mukhopadhyay
\& Ghosh\cite{mg03}, and are beyond of present scope.

The formation  of shock  in accretion disks around the compact
object was discussed by several  independent  groups (e.g.
see references\cite{msc96,yk95,ly97}). These shocks can be formed in an
accretion disk if a certain set of conditions are satisfied
simultaneously, which may happen for a specific parameter regimes
of the sub-Kepleian disk\cite{msc96,m02}.

Solving the set of disk equations\cite{mg03} and following
earlier works\cite{msc96,m03}, here we argue that when the shock
is formed in an accretion disk, the cooling ($t_{\rm cool}$) and
advection ($t_{\rm adv}$) time scale of matter from the shock
location  to the surface of a compact object are responsible for
the oscillatory  behaviour of shock that  is related to  the QPO.
If $t_{\rm cool}\sim t_{\rm  adv}$, the corresponding upper kHz
QPO frequency can  be found as
\begin{eqnarray}
\frac{1}{\nu_h}=t_{\rm adv}=\int_{x_s}^{x_{\rm in}}\frac{dx}{v},
\label{ts}
\end{eqnarray}
where $x_s$ and $x_{\rm in}$ are the location of shock and the
radius of the star respectively.

On the other hand, the lower QPO frequency may  arise due to the
Keplerian motion of the accreting fluid\cite{to99}. According to
the pseudo-Newtonian potential\cite{mm03}, the Keplerian angular
frequency of the accretion flow is given as
\begin{eqnarray}
2\pi\nu_K=\Omega_K=\frac{1}{x^{3/2}}\left[1-\left(\frac{x_{\rm ms}}{x}\right)+
\left(\frac{x_{\rm ms}}{x}\right)^2\right]^{1/2},
\label{lbo}
\end{eqnarray}
where $x_{\rm ms}$  is the radius of marginally  stable orbit
related to the specific angular momentum ($J$) of the compact
object\cite{mm03}. According to our model, this $\nu_K$ is the
lower kHz QPO frequency that varies  with the angular  frequency
of the compact  object. If we know  the angular frequency  of the
compact object and  the Keplerian radius of the  corresponding
accretion disk, this can be calculated.

As an example, Fig. 1 shows the variation of accretion speed for various
parameter regimes around 4U~1636$-$53 which has been identified as one of the
fast rotating compact star showing a pair of kHz QPO.
The outer shock locations (if there are two shocks) are usually
responsible for higher QPO frequency.
In Table 1 we enlist all our results which clearly show a very
good agreement with observations. A detailed discussion of a
similar table is given elsewhere\cite{m03}.

\begin{table*}[htbp]
\center{\large Table 1}
\begin{center}
\begin{small}
\begin{tabular}{|c|c|c|c|c|c|c|c|c|c|c|}
\hline
\hline
 source & $\alpha$ & $r_s$ & $r_k$ & $J$ & $M$  & $R$ & $\nu_h$ & $\nu_K$ &
\multicolumn{2}{c|}{observed} \\
\cline{10-11}
& & (km) & (km) & & $(M_\odot)$ & (km) & Hz & Hz & HF(Hz) & LF(Hz) \\
\hline
  & 0.05 & 15.37 & 18 & 0.2877 & 1.18 & 7.114 & 1030.8 & 719.7 & & \\
\cline{2-10}
4U~1636$-$53  &0.02 & 14.27 & 19 & 0.2877 & 1.32 & 7.23 & 1019.2 & 705.2 & 1030 & 700 \\
\cline{2-10}
& 0 & 13 & 17 & 0.2877 & 0.991 & 6.828 & 1005.2 & 715.2 & &\\
\hline
 & 0.05 &  15.37 & 15.2 & 0.2585 & 1.106 & 7.013 & 1155.4 & 907.6 & & \\
\cline{2-10}
KS~1731$-$260  &0.02 &  14.27 & 16 & 0.2585 & 1.23 & 7.16 & 1174 & 892.6 & 1159 & 898 \\
\cline{2-10}
& 0 & 13 & 14.5 & 0.2585 & 0.893 & 6.64 & 1151.9 & 863.1 & &\\
\hline
\hline
\end{tabular}
\end{small}
\end{center}
\label{table}
\end{table*}
\begin{figure}
\includegraphics[height=.35\textheight]{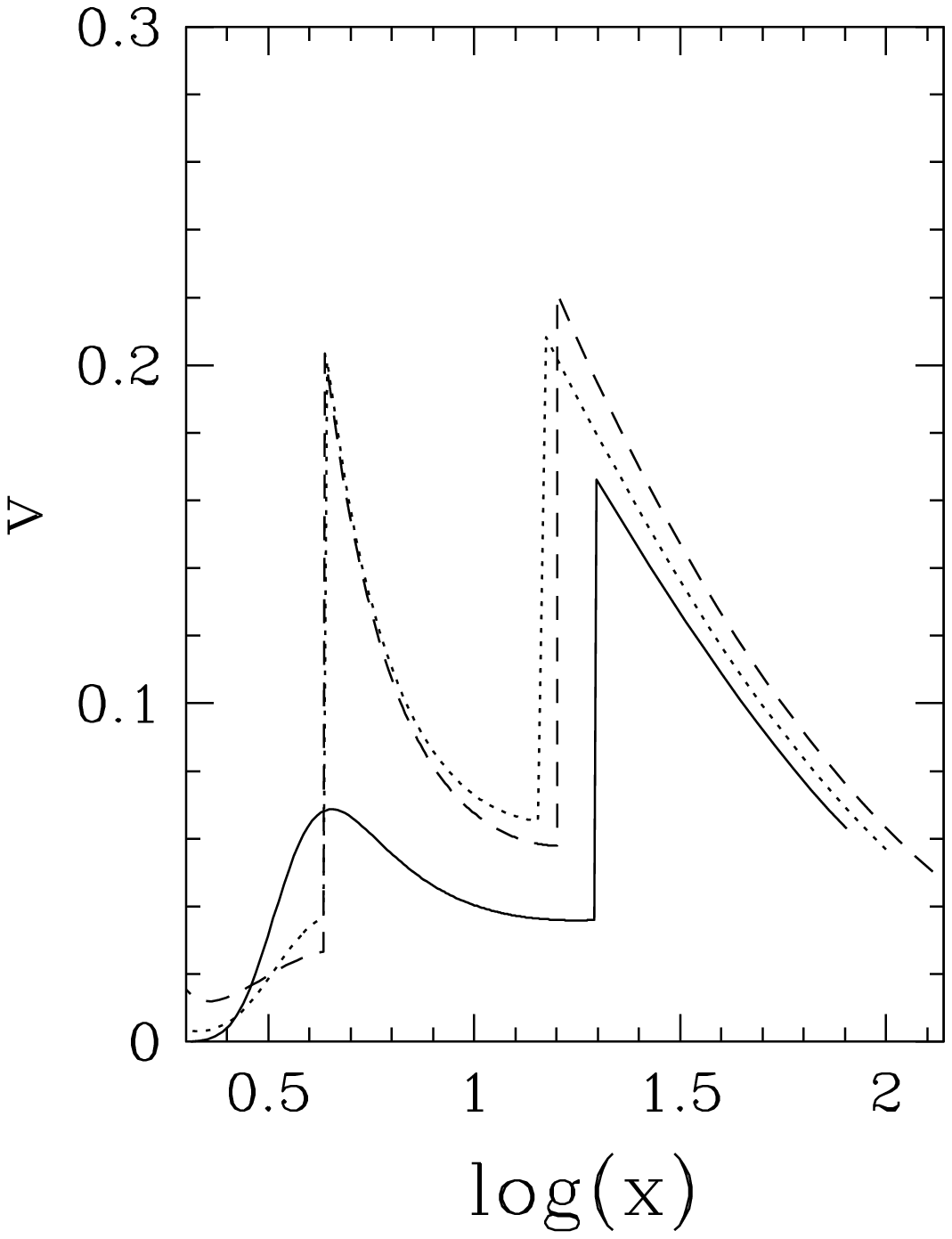}
\label{fig}
\end{figure}
\vskip-7.5cm
\hskip3.5cm
\noindent{\small Fig. 1:
Variation of accretion speed (in unit of light speed) as}

\hskip3.5cm
\noindent{\small
a function of radial coordinate (in unit of $GM/c^2$) around}

\hskip3.5cm
\noindent{\small
4U 1636-53. Solid, dotted and dashed curves are respectively }

\hskip3.5cm
\noindent{\small for inviscid flow, and two different viscous flows with viscosity,}

\hskip3.5cm \noindent{\small $\alpha=0.02$ and $\alpha=0.05$.
Here, $J=0.2877$.}
\vskip1.55cm It is perhaps for the  first time such a theoretical
calculation is presented which matches with the observation to
such a satisfactory  limit. Also the model can be checked with
the observations for other candidates. A very interesting outcome
is that the mass and radius of the compact star, supplied in our
calculation to match with observations, are not in accordance
with the conventional neutron star equation of state, but that of
a strange star\cite{d98}. Therefore this may also justify the
existence of strange stars.








%
%
%
%

\end{document}